\documentclass[apj]{emulateapj}

\newcommand{\kms}{\ifmmode{~{\rm km~s^{-1}}}\else{~km s$^{-1}$}\fi}
\newcommand{\cubecm}{\ifmmode{~{\rm cm^{-3}}}\else{~cm$^{-3}$}\fi}
\newcommand{\lsim}{\lower0.3em\hbox{$\,\buildrel <\over\sim\,$}}
\newcommand{\gsim}{\lower0.3em\hbox{$\,\buildrel >\over\sim\,$}}

\newcommand{\flux}{erg s$^{-1}$ cm$^{-2}$ Hz$^{-1}$}
\newcommand{\emis}{erg s$^{-1}$ cm$^{-2}$ Hz$^{-1}$ sr$^{-1}$}

\newcommand{\enzo}{{\sl Enzo}}
\newcommand{\Ms}{\ifmmode{M_\odot}\else{$M_\odot$}\fi}

\newcommand{\hh}{H$_2$}
\newcommand{\Ol}{$\Omega_\Lambda$}
\newcommand{\Om}{$\Omega_M$}
\newcommand{\Ob}{$\Omega_b$}

\newcommand{\tvir}{\ifmmode{T_{\rm{vir}}}\else{$T_{\rm{vir}}$}\fi}
\newcommand{\mvir}{\ifmmode{M_{\rm{vir}}}\else{$M_{\rm{vir}}$}\fi}

\newcommand{\rr}{$r_{200}$}
\newcommand{\lya}{Ly$\alpha$}
\newcommand{\jj}{\ifmmode{J_{21}}\else{$J_{21}$}\fi}
\newcommand{\juv}{\ifmmode{J_{21}^{912}}\else{$J_{21}^{912}$}\fi}
\newcommand{\jlw}{\ifmmode{J_{21}^{\rm{LW}}}\else{$J_{21}^{\rm{LW}}$}\fi}
\newcommand{\flw}{\ifmmode{F_{\rm{LW}}}\else{$F_{\rm{LW}}$}\fi}

\begin{document}

\shorttitle{SUPPRESSION OF H$_2$ COOLING IN THE UVB}
\shortauthors{WISE \& ABEL}

\title{Suppression of H$_2$ Cooling in the Ultraviolet Background}
\author{John H. Wise\altaffilmark{1,2} and Tom Abel\altaffilmark{1}}
\altaffiltext{1}{Kavli Institute for Particle Astrophysics and Cosmology,
  Stanford University, 2575 Sand Hill Road, MS 29, Menlo
  Park, CA 94025}
\altaffiltext{2}{Laboratory for Observational Cosmology, NASA Goddard
  Space Flight Center, Greenbelt, MD 20771}
\email{jwise, tabel@slac.stanford.edu}

\begin{abstract}

  The first luminous objects in the concordance cosmology form by
  molecular hydrogen cooling in dark matter dominated halos of masses
  $\sim 10^6 \Ms$. We use Eulerian adaptive mesh refinement
  simulations to demonstrate that in the presence of a large soft
  ultraviolet radiation background, molecular hydrogen is the dominant
  coolant.  Even for very large radiation backgrounds, the halo masses
  that cool and collapse are up to two orders of magnitude smaller
  than the halos that cool via atomic hydrogen line cooling. The
  abundance of cooling halos and the cosmic mass fraction contained
  within them depends exponentially on this critical mass scale.
  Consequently, the majority of current models of cosmological
  reionization, chemical evolution, supermassive black hole formation,
  and galaxy formation underestimate the number of star forming
  progenitors of a given system by orders of magnitude.  At the
  highest redshifts, this disagreement is largest. We also show that
  even in the absence of residual electrons, collisional ionization in
  central shocks create a sufficient amount of electrons to form
  molecular hydrogen and cool the gas in halos of virial temperatures
  far below the atomic cooling limit.

\end{abstract}


\section{MOTIVATION}

Cosmic structure forms hierarchically.  Any object in the universe
today, started with copious numbers of small progenitors at redshifts
currently inaccessible to direct observations.  Traditionally in
galaxy formation \citep{Rees77, White78, Dekel87, White91, Baugh03}
\tvir~= $10^4$ K halos are assumed to be the first cooling
halos. Nevertheless since the late 1960's it has been known that
molecular hydrogen, formed in the gas phase, can dominate cooling in
objects of smaller virial temperature and mass \citep{Saslaw67,
  Peebles68b, Yoneyama72, Haiman96, Tegmark97, Abel98,
  Abel00}. Neglecting this early phase of \hh~cooling halos has been
justified by arguing that \hh~is destroyed via radiative feedback
effects \citep[cf.][]{Dekel87, Haiman97b, Haiman00, Glover01,
  Bromm03}. The photo-dissociation of \hh~via the Solomon process by
an early soft ultraviolet background (UVB) is generally assumed as the
main reason \citep{Oh02, Ciardi05, Haiman06}.

%
%
\begin{figure}[!b]
\begin{center}
\resizebox{\columnwidth}{!}{\includegraphics*{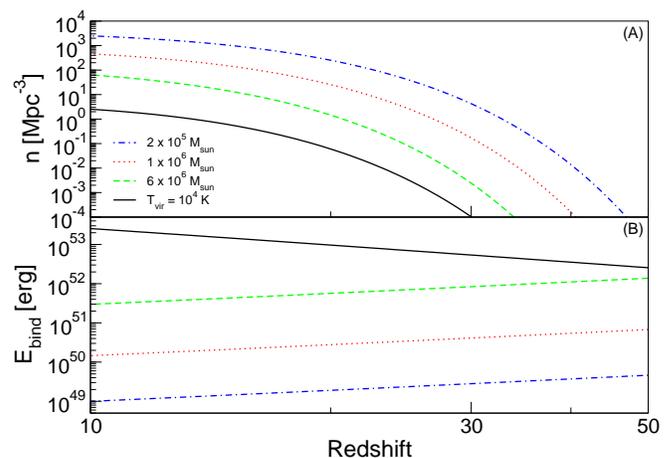}}
\caption{\label{fig:ps} \textit{Panel A}: Sheth Tormen number density
  of dark matter halos as a function of redshift for \tvir~= $10^4$ K,
  $M = 2 \times 10^5, 10^6,~\rm{and}~6 \times 10^6~\Ms$ using WMAP 3
  year data for parameters. \textit{Panel B}: Binding energies as a
  function of redshift for the corresponding halos (same line styles
  as in panel A)}
\end{center}
\end{figure}

The mass scale of halos considered enters exponentially in the
collapsed mass fraction and the abundance of halos.  Figure
\ref{fig:ps} shows the predicted abundances of the earliest building
blocks of galaxy formation as a function of redshift for the latest
concordance cosmology using the Sheth-Tormen formalism \citep{Press74,
  Sheth02}. The different lines correspond to different virial
masses. The solid line corresponds to halos with virial temperatures
of $10^4$ K, the temperature at and above which atomic hydrogen line
cooling is dominant. At redshift 30, e.g., the difference of
abundances of $2 \times 10^5$ \Ms~and \tvir~= $10^4$ K halos is five
orders of magnitude. Even at redshift 10 this disparity is still a
factor of a thousand. When studying reionization and chemical
evolution of galaxies and the intergalactic medium, one needs to
consider stellar feedback. The simple fact that the binding energy of
the gas of smaller mass halos is even less than the kinetic energy
deposited by even one supernova (SN) is illustrated in Figure
\ref{fig:ps}B. Surely whether the atomic hydrogen line (\lya) cooling
halos are formed from pristine primordial gas or are mergers of many
tens of progenitors that massive stars have enriched and expelled the
gas from should make a significant change in their further evolution.
The minimum mass of star forming halos is undoubtedly an important
issue independent of the techniques employed to study structure
formation.

Advances in cosmological hydrodynamics and its numerical methods
\citep{Cen92a, Zhang95, Katz96, Abel97, Anninos97, Bryan98, Gnedin01,
  Ricotti02a, Ricotti02b} allow now detailed investigations of all the
relevant physical processes. Modeling the expected negative feedback
from an early soft UVB is straightforward as a background flux only
causes a spatially constant photo-dissociation rate in the chemical
reaction network being solved when \hh~does not exist at high enough
abundances to self-shield. \citet[][MBA01 hereafter] {Machacek01} used
Eulerian adaptive mesh refinement (AMR) simulations to investigate the
role of such a \hh~dissociating (Lyman-Werner; LW) background on the
minimum mass of halos within which primordial gas can first cool for a
variety of radiation amplitudes.  In addition to a LW background, the
collapse of halos within relic \ion{H}{2} regions can be either
delayed or catalyzed.  \citet{Mesigner06} used AMR simulations with a
short-lived 3 Myr hydrogen ionizing UVB that simulates a nearby
massive, metal-free (Pop III) star.  They found that halo collapses
are prolonged if \juv~$\gsim$ 0.1 and catalyzed if below this critical
value, where \juv~is in units of $10^{-21}$ \emis~at a wavelength of
912\AA.  In the case of a large UVB, the collapse is delayed due to
lower gas densities and higher cooling times.  In the small UVB
regime, excess free electrons in the relic \ion{H}{2} region
accelerate \hh~formation.  In both cases, feedback in relic \ion{H}{2}
subsides after $\sim$30\% of a Hubble time.  Strong suppression of
\hh~formation also occurs in $10^6 \Ms$ halos with a LW background
\jlw~$>$ 0.01.  \citet[][YAHS03 hereafter] {Yoshida03} similarly
addressed this issue using smoothed particle hydrodynamics (SPH). They
found an additional effect on the minimum collapse mass of dynamical
heating from the mass accretion history of the halo. As the heat input
increases, the virial temperature must rise before \hh~cooling can
start to dominate, and a cool phase develops in the center of the
potential well.

Self-consistent calculations in which the sources produce the
radiation backgrounds which in turn affect the number of new sources
are feasible so far only with semi-analytic approaches \citep[][WA05
hereafter] {Haiman00, Wise05} and small volume cosmological
simulations at low spatial resolutions \citep{Ricotti02a,
  Ricotti02b}. From these studies, one can derive realistic upper
limits on the amplitude of the expected soft UVB. In all studies that
include radiation sources in halos less than $10^4$ K halos, the
largest the soft UVB flux can get before the $T > 10^4 K$ halos
dominate the emission is $\jlw \sim 1$ \citep [cf.][WA05]{Haiman00,
  Ricotti02a, Ricotti02b}. Interestingly, for a LW intensity of $\jlw
\sim 0.1$, MBA01 found that $2 \times 10^6 \Ms$ halos were still able
to cool and collapse.  On the other hand at that \jlw, YAHS03 suggest
negative feedback should become so strong that the critical
\hh~fraction for cooling cannot be reached and cooling will not occur.
However, they did not explore this further with detailed higher
resolution simulations to check whether their analytical expectation
would hold.

We present a series of fourteen very high resolution Eulerian AMR
simulations designed to see how the largest possible feedback may
raise the minimum mass in which primordial gas will cool by molecular
hydrogen. The simulations techniques and details of the suite of
calculations is the topic of the next section. In the following
sections, we describe the results that show \hh~cooling cannot be
neglected in early structure formation. In the discussion, we describe
the nature of the UVB and why \hh~cooling can occur in such large
radiation backgrounds.  We also comment on the large range of
questions in cosmological structure formation that this conclusion
affects.

%
%
\begin{deluxetable}{lccccc}
\tabletypesize{}
\tablewidth{\columnwidth}
\tablecaption{Simulation Properties\label{tab:sims}}

\tablehead{ \colhead{Name} & \colhead{\hh} & \colhead{Residual e$^-$}
  & \colhead{\flw} & \colhead{z$_a$} & \colhead{z$_b$}
}
\startdata
H2         \dotfill& Yes & Yes & 0         & 29.7  & 31.1 \\
H2LW22     \dotfill& Yes & Yes & 10$^{-22}$ & 28.3  & 27.5 \\
H2LW21     \dotfill& Yes & Yes & 10$^{-21}$ & 24.4  & 24.7 \\
H2LW20     \dotfill& Yes & Yes & 10$^{-20}$ & 20.5 & 22.4 \\
noe-H2     \dotfill& Yes & No  & 0         & 18.7  & 23.4 \\
noe-H2LW20 \dotfill& Yes & No  & 10$^{-20}$ & 16.8  & 21.4 \\
H+He       \dotfill& No  & Yes & 0         & 15.9  & 16.8
\enddata
\tablecomments{These simulations are performed for both realizations.}
\end{deluxetable}

%
%
\begin{figure*}[t]
\begin{center}
\resizebox{0.8\textwidth}{!}{\includegraphics*{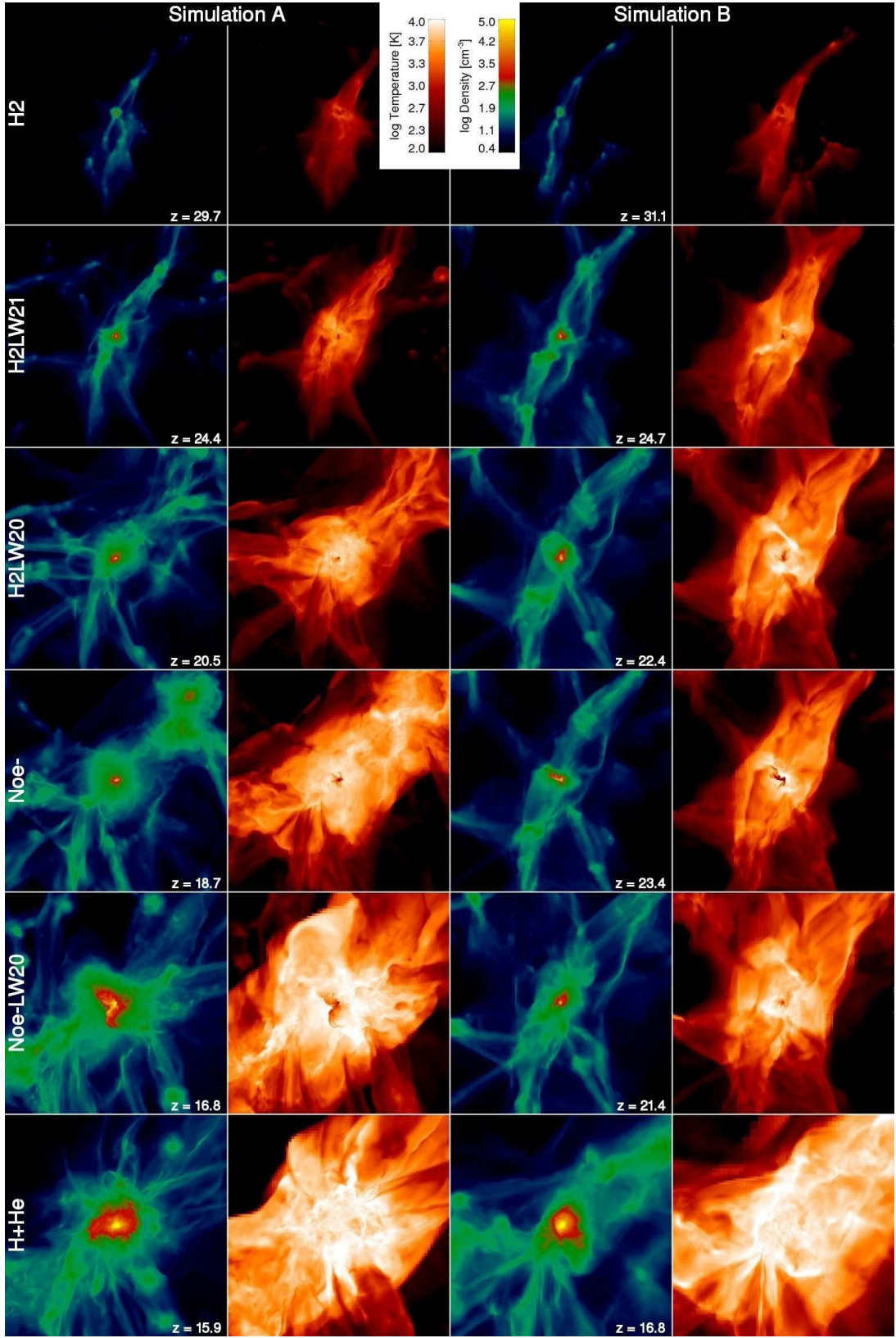}}
\caption{\label{fig:rho} Density-squared weighted projections in
  simulation A (left two columns) and B (right two columns) of the gas
  density (first and third columns) and temperature (second and fourth
  columns) at the times when the most massive halo starts to cool and
  collapse above an overdensity of $10^7$ in the models. The rows show
  the H2, H2LW21, H2LW20, and noe-H2, noe-H2LW20, and H+He runs from
  top to bottom, respectively.  Note the complex structure for the
  SimA-Noe-LW20 and SimB-Noe-H2 run in which central shocks lead to
  the formation of free electrons that promote the formation of H2 and
  triggering the collapse.  The field of view in all panels is 1.2
  proper kpc.  The color maps are equal for all images.}
\end{center}
\end{figure*}

\section{SIMULATIONS AND ASSUMPTIONS}

We use the Eulerian AMR hydrodynamic code \enzo~\citep{Bryan97,
  Bryan99} to study the importance of \hh~cooling in early galaxy
formation.  \enzo~uses an $n$-body adaptive particle-mesh solver
\citep{Couchman91} to follow the dark matter (DM) dynamics.  We
perform two cosmological realizations with different box sizes and
random phases and WMAP 1 year parameters of ($h$, \Ol, \Om, \Ob,
$\sigma_8$, $n$) = (0.72, 0.73, 0.27, 0.024$h^{-2}$, 0.9, 1)
\citep{Spergel03}.  The significantly different third year WMAP
\citep[WMAP3;][] {Spergel07} results favor lesser small-scale power
that delays high-redshift structure formation by $\sim$40\% and alters
the statistical properties of DM halos \citep{Alvarez06}.  The ratio
\Om/\Ob~also only lowered by 5\% to 5.70.  However these differences
have no effect on the evolution and assembly of individual halos
studied here that have typical mass accretion histories.

The initial conditions are the same as in \citet{Wise07}.  Both
realizations have a top grid with a resolution of 128$^3$ with three
nested subgrids with twice finer resolution and are initialized at z =
129 (119)\footnote{To simplify the discussion, simulation A will
  always be quoted first with the value from simulation B in
  parentheses.} with the COSMICS package \citep{Bertschinger95,
  Bertschinger01}.  The box size is 1.0 (1.5) comoving Mpc.  The
innermost grid has an effective resolution of 1024$^3$ with DM
particle masses of 30 (101) \Ms and a side length of 250 (300)
comoving kpc.  We refine the AMR grids when either the DM (gas)
exceeds three times the mean DM (gas) density on the same level.  We
also refine so that the local Jeans length is resolved by at least 4
cells.

We focus on the region containing the most massive halo in the
simulation box and follow its evolution until it collapses to an
overdensity of $10^7$ that corresponds to a refinement level of 15 and
a spatial resolution of $\sim$3000 (4000) proper AU.

We perform each realization with seven sets of assumptions.  Table
\ref{tab:sims} summarizes them.  We use a nine-species (H, H$^{\rm
  +}$, He, He$^{\rm +}$, He$^{\rm ++}$, e$^{\rm -}$, H$_2$, H$_2^{\rm
  +}$, H$^{\rm -}$) non-equilibrium chemistry model \citep{Abel97,
  Anninos97} for all runs except the H+He runs that do not include
\hh~cooling.  The nine-species runs are specified by ``H2'' and use
the \hh~cooling rates from \citet{Galli98}.  Above number densities of
$10^4 \cubecm$ or in an intense ultraviolet radiation field, the
excited states of \hh~become populated.  The \hh~collisional
dissociation rates from \citet{Abel97} are calculated in the ground
state; therefore we use a density dependent \hh~dissociation rate from
\citet{Martin96} that considers this phenomenon.  Runs with
\hh~dissociating (Lyman-Werner; LW) radiation are denoted by ``LW''
followed by its negative log-flux.  We set \flw~to $10^{-22}$,
$10^{-21}$, and $10^{-20}$ \flux~because the first two are typical
values one finds in semi-analytic models of reionization and the
latter investigates the case of an very large UVB \citep[e.g.][]
{Haiman00, Wise05}.  We use the \hh~photo-dissociation rate
coefficient for the Solomon process from \citet{Abel97} of
$k_{\rm{diss}} = 1.1 \times 10^8 \flw$ s$^{-1}$.  We do not consider
the self-shielding of LW photons.  Because the molecular core only
becomes optically thick in the late stages of collapse and above
column densities of 10$^{14}$ cm$^{-2}$ \citep{Draine96}, we expect
our results to not be drastically affected by neglecting LW
self-shielding.  Additionally, LW self-shielding may be unimportant up
to column densities of $10^{20} - 10^{21}$ cm$^{-2}$ if the medium
contains very large velocity gradients and anisotropies
\citep{Glover01}.

Free electrons are necessary to form \hh~in the gas phase.  In order
to restrict \hh~formation to \lya~line cooling halos in our ``noe-''
calculations, we reduce the residual free electron fraction from
$\sim10^{-4}$ \citep{Peebles68a, Shapiro94} to a physically low
$10^{-12}$ at the initial redshift.  This setup is designed to find
the first halos that can collapse and form stars once free electrons
from collisionally ionized hydrogen becomes available to catalyze
\hh~formation \citep{Shapiro87}.

This work is an extension of the original work of MBA01, adding the
calculations with \flw~= $10^{-20}$ and ones in which \hh~cannot cool
until \lya~cooling becomes efficient.  We consider these extreme cases
to strengthen the point made in MBA01 in which a UVB only increases
the critical halo collapse mass, never completely suppressing the
crucial importance of \hh~formation and cooling.  Our maximum spatial
resolution in the finest AMR level is a factor of four smaller than
MBA01; however, this does not cause any differences between our work
and MBA01 because these finest grid patches only exist in the dense,
central core during the final 150 kyr of the collapse.

%
%
\begin{figure}[t]
\begin{center}
\resizebox{\columnwidth}{!}{\includegraphics*{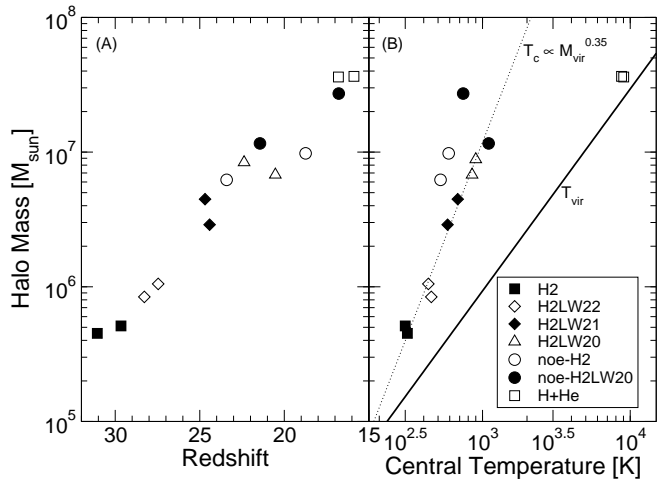}}
\caption{\label{fig:mass} \textit{Panel A}: Halo masses of the most
  massive halos as function of redshift when they reach a central
  overdensity of $10^7$. This allows to translate the mass values in
  Panel B to be converted to cooling redshifts.  It marks the runs
  with H+He (\textit{filled squares}), \hh~(\textit{open squares}),
  $\flw=10^{-22}$ (\textit{open diamonds}), $\flw=10^{-21}$
  (\textit{filled diamonds}), no residual electrons (\textit{open
    circles}), and extreme feedback noe-LW20 (\textit{filled circles})
  runs.  The two data points for each symbol represent simulations A
  and B.  Even for the most extreme cases of feedback cooling occurs
  much earlier than in the atomic line cooling only
  case. \textit{Panel B}: Central temperatures of most massive halo in
  the simulation as a function of its mass at different redshifts.
  The virial temperature computed form the dark matter halo mass at
  redshift 20 is the solid line.  The dotted line is the fitted
  relationship between the central gas temperatures and the halo mass
  in models with residual electrons and \hh~cooling.}
\end{center}
\end{figure}

\subsection{Virial Temperature}
\label{sec:tvir}

In galaxy formation models, the virial temperature is a key quantity
as it controls the cooling and star formation rates in a given halo.
We define a halo as the material contained in a sphere of radius
\rr~enclosing an average DM overdensity $\Delta_c$ of 200.  For an
isothermal singular sphere, the virial temperature
\begin{equation}
\label{eqn:tvir}
T_{vir} = \frac{\mu m_p V_c^2}{2k},
\end{equation}
where $V_c^2 = GM/r_{200}$ is the circular velocity \citep[see][with
$\beta$ = 1]{Bryan98}.  Here $\mu$ is the mean molecular weight in
units of the proton mass $m_p$, and $k$ is Boltzmann's constant.  We
use this definition of \tvir~in this paper with $\mu$ = 0.59.  We
choose this value of $\mu$ to be consistent with the literature on
galaxy formation even though the halos presented in this paper are
neutral and have $\mu$ = 1.22.

\section{RESULTS}

We first describe the halo properties at collapse.  Then we compare
them to previous studies of collapsing halos in the presence of a soft
UVB.

\subsection{Halo Properties}

Figure \ref{fig:rho} shows density-squared weighted projections of gas
density and temperature when each calculation can cool and collapse to
an overdensity of 10$^7$.  It illustrates the large difference in the
sizes and morphologies of the collapsing halos in the various cases of
negative feedback.  All panels have the same field of view of 1.2
proper kpc and same color scales.  It is clear from the relative sizes
of the collapsing halos that the critical halo mass to cool increases
with the amount of negative feedback.  The virial shock and numerous
central shocks heat the gas to the virial temperature.  The central
shocks create fine structure seen in the temperature projections.  In
all of the \hh~cases, we see neither fragmentation nor large-scale
disk formation.  The internal structures of the halos with \hh~cooling
and residual free electrons are similar to previous studies of Pop III
star forming halos \citep[MBA01;][]{Abel00, Abel02, Bromm02,
  Yoshida03}, exhibiting a turbulent medium with a radially
monotonically decreasing density profile and a cool central core.

Figure \ref{fig:mass}A depicts the halo mass and redshift when the
halo collapses for all of the runs, and Figure \ref{fig:mass}B shows
their central temperature at the same epoch.  The collapse redshifts,
$z_a$ and $z_b$, are also listed in Table \ref{tab:sims} for
simulations A and B, resepectively.  As seen in other studies (MBA01,
YAHS03), the minimum DM halo mass to collapse increases with the
background intensity.  The H+He case predictably collapses at
\tvir~$\sim 10^4$ K, and all of the halos with \hh~cooling collapse at
much smaller masses.  The temperature of the central core increases
with halo mass from 300 K to 1000 K for halo masses $4 \times 10^5
\Ms$ and $10^7 \Ms$.  Restricting the data to models with residual
electrons, the central temperature increases as a power-law, 
\begin{equation}
\label{eqn:Tc}
T_c = A \mvir^B, 
\end{equation}
where $A = 3.1^{+1.3}_{-0.9}$, $B = 0.355 \pm 0.024$, and
\mvir~is in units of solar masses.  This relationship is plotted in
Figure \ref{fig:mass}.

\begin{figure}[t]
\begin{center}
\resizebox{\columnwidth}{!}{\includegraphics*{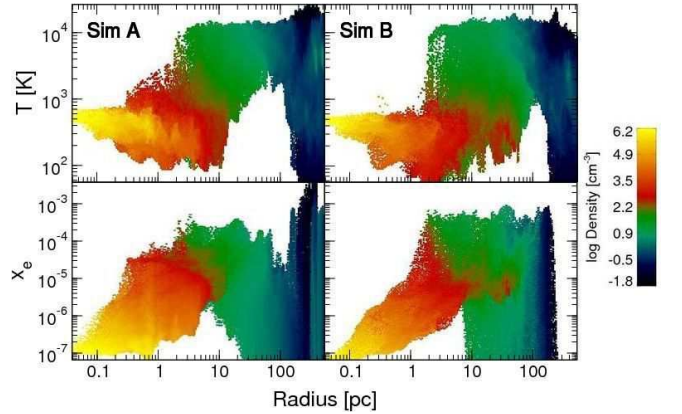}}
\caption{\label{fig:efrac} Radial profiles of temperature
  (\textit{top}) and electron fraction (\textit{bottom}) colored by
  density for the ``noe-'' simulations with no residual free electrons
  or UVB in simulation A (\textit{left}) and B (\textit{right}).  The
  virial temperatures of these halos are 4600K and 4200K for
  simulation A and B, using equation (\ref{eqn:tvir}) with $\mu$ =
  0.59.}
\end{center}
\end{figure}

With neither residual electrons nor an UVB (noe-H2), the most massive
halo collapses at $9.8 \; (6.2) \times 10^6 \Ms$ at z = 18.7 (23.4).
Here \hh~formation in the gas phase can only become important when
sufficient free electrons are created by collisional ionization.
Virial heating in the center of halos can increase temperatures up to
twice the virial temperature \citep{Wise07} that collisionally ionizes
hydrogen in the central shocks and initiates \hh~cooling
\citep{Shapiro87} in halos well below virial temperatures of 10$^4$ K.
These shocks are abundant throughout the central regions.  Figure
\ref{fig:efrac} shows radial profiles of temperature and electron
fraction for both simulations and depicts gas shock-heating up to $2
\times 10^4$ K and raising electron fractions up to 10$^{-3}$.  The
electron fractions remain at unrealistically low values less than
10$^{-6}$ in low density regions where gas has not been collisionally
ionized.  The higher density regions have condensed to densities above
$3 \times 10^2 \cubecm$ after free electrons in protogalactic shocks
induced \hh~cooling.

A similar but extreme model, noe-H2LW20, demonstrates that even in the
presence of a very large UVB of \flw~= 10$^{-20}$ gas is able to form
a cool and dense central molecular core at a mass of $2.7 \; (1.1)
\times 10^7 \Ms$ at redshift 16.8 (21.4).  Two major mergers in
simulation A occur between z = 17--21, and the associated heating
allows the halo to begin cooling by \hh.  A central core only forms
once the system is adequately relaxed after the mergers, which causes
the collapse mass difference between the realizations.

By not fully resolving weak shocks in our main calculations, it is
possible to underestimate the electron fraction.  We performed
SimB-H2LW20 with an additional refinement criterion that resolves the
``cooling length'', $l_{\rm{cool}} = t_{\rm{cool}} / c_s$, by at least
2 cells.  The large- and small-scale structure in the simulation is
unchanged.  When we resolve these weak shocks, the increased electron
fraction marginally accelerates the collapse, which occurs 780 kyr
earlier at z = 22.5.  The virial mass at this time is $8.0 \times 10^6
\Ms$ compared with $8.4 \times 10^6 \Ms$.  Hence we believe that the
critical halo mass to collapse as a function of the LW background is
independent of this refinement criterion.

The combination of a recent major merger and collisional ionization
produces complex structures as seen in the density and temperature
projections of the SimA-Noe-LW20 and SimB-Noe- calculations in Figure
\ref{fig:rho}, unlike the other H2 models with a single cool central
core.

%
%

\vspace{1em}
\subsection{Comparison to Previous Studies}

Through a series of AMR calculations with varying UVB intensities,
MBA01 found the minimum DM halo mass 
\begin{equation}
  \label{eqn:mass_crit}
  M_{\rm{crit}} = 2.5 \times 10^5 + 1.7 \times 10^6 (\flw/10^{-21})^{0.47} \; \Ms  
\end{equation}
in order to cool and
condense 4\% of the baryons.  This fraction of cool and dense gas
agrees with simulations of the formation of Pop III stars
\citep{Abel02, Yoshida06}.  There is some scatter of $\sim$0.5 dex in
this threshold mass (see also YAHS03).  For the UVB intensities used
in our models (\flw~= 0, 10$^{-22}$, 10$^{-21}$, 10$^{-20}$), the
critical collapse masses are $2.5 \times 10^5$, $8.4 \times 10^5$,
$2.0 \times 10^6$, and $5.4 \times 10^6$\Ms.  Our calculations with
\hh~cooling and residual free electrons agree with the results of
MBA01.

YAHS03 studied the minimum collapse mass but also included the effects
of self-shielding.  Through their SPH simulations and arguments using
equilibrium \hh~abundances, they conclude that an UVB intensity of
\jj~= 0.1 nearly prevents halo collapses below \tvir~$\simeq$ 7000 K
where \lya~cooling becomes efficient.  They also deduce that \jj~= 1.0
completely prevents any \hh~cooling in these low-mass halos, based on
\hh~dissociation timescales.  We find the contrary in our H2LW21 and
H2LW20 calculations where the most massive halo collapses with a mass
of $4.5\;(2.9) \times 10^6 \Ms$ and $8.4\;(6.8) \times 10^6 \Ms$,
respectively.  Even in our noe- runs, the halo collapses when
\tvir~$\sim$ 4000 K, i.e. before \lya~cooling becomes important, which
is around the same mass scale that the H2LW20 runs condense.  We
ignore self-shielding in our calculations, but this would only
decrease the critical collapse mass and strengthens our main
conclusion that \hh~cooling is always dominant, even in the presence
of a large LW flux.


YAHS03 used cosmological SPH simulations and \hh~formation and
dissociation timescales to argue that a LW background intensity of
$\jlw > 0.1$ suppresses \hh~formation so halos cannot cool before
virial temperatures of 7000 K are reached.  Employing the same
argument, we see that the \hh~formation timescale
\begin{equation}
  \label{eqn:h2form}
  t_{\rm{H_2}} = \frac{n_{\rm{H_2}}}{k_{\rm{H^-}} n_{\rm{H}} n_{\rm{e}}}
  = \frac{f_{\rm{H_2}}}{0.92 k_{\rm{H^-}} f_{\rm{e}} n}
  \approx 30 \; \textrm{kyr} ,
\end{equation}
with typical central values found in high-redshift halos before any
radiative cooling becomes efficient \citep[see][]{Wise07}.  Here
$f_{\rm{H_2}} = 10^{-6}$ and $f_{\rm{e}} = 10^{-4}$ are the \hh~and
electron number fraction, respectively, and $n = 10\cubecm$ is the
baryon number density.  $k_{\rm{H^-}} \approx 10^{-15}$ cm$^3$
s$^{-1}$ is the H$^-$ formation rate coefficient by electron
photo-attachment at $T = 1000$ K \citep{Abel97}.  This timescale is a
factor of 1000 smaller than the value calculated in YAHS03 because we
use the quantities from the halo center as compared to the mean
values.  The \hh~dissociation timescale is $k_{diss}^{-1}$ =
23/\jj~kyr, which is comparable with $t_{\rm{H_2}}$ using the values
above.

The halo characteristics and the collapse redshift will likely depend
on halo merger histories as seen in these two realizations.  The
better statistics of MBA01 sampled this effect well.  Here the scatter
of threshold mass is $\sim$0.5 dex and is smaller than the mass
difference between halos with virial temperatures of 4000K and 10000K.
Thus our limited sample of halos should not change our result of the
importance of \hh~cooling in halos well below \tvir~= 10$^4$ K, even
with very large LW radiation backgrounds.

\section{DISCUSSION}

%
%
\begin{figure*}[t]
\begin{center}
\resizebox{0.8\textwidth}{!}{\includegraphics*{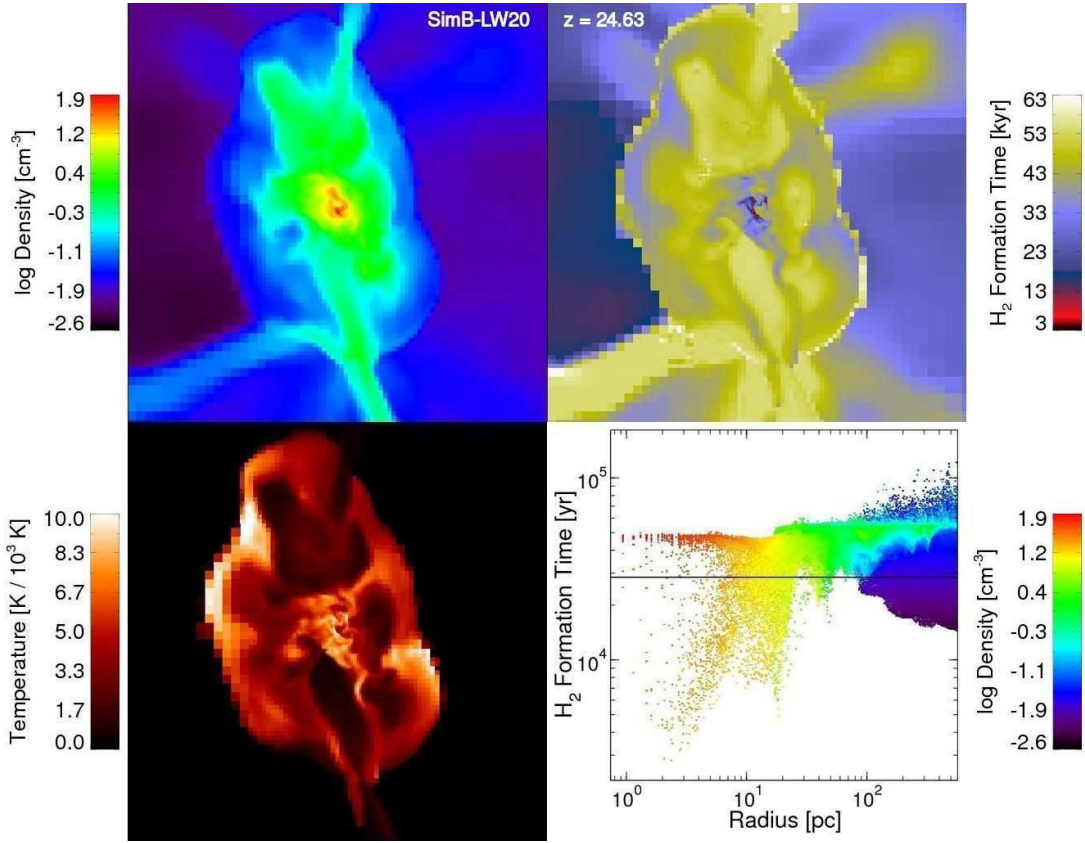}}
\caption{\label{fig:h2form} The most massive halo in SimB-LW20 (\flw~=
  $10^{-20}$) twenty million years before the cooling core collapses.
  Slices of gas density (\textit{left}) and the \hh~formation
  timescale (Eq. [\ref{eqn:h2form}]; \textit{right}) through the
  densest point in the halo are in the \textit{top} row.  The
  \textit{bottom} row contains a slice of temperature (\textit{left})
  and a radial profile of the \hh~formation timescale, colored by gas
  density.  The phase diagram and density slice have the same color
  scale.  The slices have a field of view of 500 proper pc.  In the
  central shocks, \hh~formation timescales are lower than the
  dissociation timescale of 28 kyr with \flw~= 10$^{-20}$ that is
  denoted by the horizontal line in the radial profile.  The central
  core is now efficiently cooling and will collapse 20 Myr after these
  data.}
\end{center}
\end{figure*}

Structure formation in the high-redshift universe is contained within
shallow potential wells that are sensitive to negative feedback from a
UVB.  Additionally local positive and negative feedback will influence
star formation and further complicate estimates of halo mass scales.
Some examples include

\begin{itemize}
\item \textit{Positive feedback}--- Enhanced \hh~formation in relic
  \ion{H}{2} regions \citep[e.g.][]{Ferrara98, OShea05, Johnson07} and
  ahead of the \ion{H}{2} ionization front \citep{Ricotti01, Ahn07},
  dust and metal line cooling \citep{Glover03, Schneider06,
    Jappsen07},
\item \textit{Negative feedback}--- Baryonic explusion from host halos
  \citep{Whalen04, Kitayama04, Yoshida07, Abel07}, photo-evaporation
  \citep{Susa06}, entropy floors \citep{Oh03}.
\end{itemize}

These processes are not within the scope of this paper and will be
considered in later publications that utilize three-dimensional
radiation hydrodynamic simulations with Pop III star formation.  Here
we only focused on the effects of a UVB on low-mass halos.

\subsection{The Nature of the UVB}

The intensity of the UVB is a monotonically increasing function of
redshift as more halos form stars.  The UVB increases on the order of
a Hubble time, which is much shorter than a dynamical time of a
collapsing halo and justifies the use of a constant intensity in our
calculations.

Self-consistent studies that evolve the UVB according to star
formation rates only find \jlw~to be in the range of 0.01 and 0.1 at
redshifts 15--20 (YAHS03, WA05).  WA05 calibrated their model against
the WMAP1 measurement of $\tau$ = 0.17.  With the WMAP3 result of the
electron scattering optical depth $\tau$ = 0.09 and less small-scale
power, UVB intensities will be even lower at these redshifts.

We can relate reionization to LW radiation by equating \jj~in the LW
band to a common quantity in reionization models, the ratio of emitted
hydrogen ionizing photons to baryons, $n_{\gamma, \rm{HI}} /
\bar{n}_b$, where $\bar{n}_b \simeq 2 \times 10^{-7} (1+z)^3 \cubecm$
is the cosmic mean of the baryon number density.  Assuming that
$J_{\rm{LW}}$ is constant in the LW band, the number density of LW
photons is
\begin{eqnarray}
  n_{\gamma, \rm{LW}} &=& \frac{4\pi}{c} \int^{\nu_2}_{\nu_1} 
  \frac{J_{\rm{LW}}}{h_p \nu} \; d\nu\nonumber \\
  &=& 1.19 \times 10^{-5} \jlw \; \textrm{cm}^{-3},
\end{eqnarray}
where $h_p$ is Planck's constant and $\nu_1, \nu_2$ = $2.70 \times
10^{15}$ Hz, $3.26 \times 10^{15}$ Hz bound the LW band.  To relate
\jj~to $n_{\gamma, \rm{HI}} / \bar{n}_b$, we must consider the
intrinsic ionizing spectrum and absorption from the IGM and host halo.
At redshift 20, the majority of star forming halos host Pop III stars
that emit a factor $\phi_{\rm{HI}} \simeq 10$ more hydrogen ionizing
photons than LW photons because of its $\sim10^5$~K surface
temperature.  Since the number density of sources exponentially
increases with redshift, the majority of the early UVB at a given
redshift originates from cosmologically nearby ($\Delta z / z \sim
0.1$) sources.  Lyman line resonances absorb a fraction $f_{\rm{abs}}
\sim 0.1$ of the LW radiation in the intergalactic medium in this
redshift range, producing a sawtooth spectrum \citep{Haiman97a}.
Additionally, absorption in the host halo reduces the number of
ionizing photons that escape into the IGM by a fraction
$f_{\rm{esc}}$.  For Pop III halos, this factor is close to unity
\citep{Yoshida07, Abel07}.  By considering these multiplicative
processes, we now estimate
\begin{eqnarray}
  \label{eqn:ph_ratio}
  \frac{n_{\gamma, \rm{HI}}}{\bar{n}_b} &=& 
  \frac{n_{\gamma, \rm{LW}}}{\bar{n}_b} \;
  \left( \frac{1+z}{20} \right)^{-3} \;
  \phi_{\rm{HI}} \; f_{\rm{esc}} \; f_{\rm{abs}}^{-1}
  \nonumber \\
  &=& 0.64 \; J_{21} \left(\frac{1+z}{20}\right)^{-3}
  \left(\frac{\phi_{\rm{HI}}}{10}\right)
  \left(\frac{f_{\rm{esc}}}{1}\right) \\
  & & \times \left(\frac{f_{\rm{abs}}}{0.1}\right)^{-1} \nonumber
\end{eqnarray}
This estimate is in agreement with the reionization models of
\citet{Haiman00} and WA05 \citep[see also][]{Gnedin97}.  These models
find that sources produce a large UVB of \jj~$\sim$ 1 prior to
reionization.  When Pop III stars dominate the UVB, the LW radiation
will be small in comparison to the volume averaged hydrogen ionizing
emissivity because of the intrinsically hard Pop III spectra that
peaks at $\sim$300\AA.  Hence high-redshift halos should not be
exposed to a large UVB, i.e. \jj~$\gsim$ 0.1, and \hh~formation will
remain important before reionization.

Nearby star formation can boost the LW radiation over its background
value, but these bursts are short-lived as Pop III lifetimes are only
$\sim$3 Myr \citep{Schaerer02}.  For example, a 100 \Ms~star produces
10$^{50}$ LW photons s$^{-1}$ and will produce $\jlw > 0.1$ in the
surrounding 3 proper kpc, neglecting any \hh~self-shielding.


The LW background is uniform outside these spheres of influence.  The
bursting nature of Pop III star formation does not affect the time
evolution of the background.  The intensity only depends on the number
of sources in a redshift range $\Delta z / z$ = 13.6 eV / 11.18 eV --
1, where the two energies bound the LW band, because any radiation
redward of the Lyman break contributes to the LW background.  Using a
conservative minimum halo mass for Pop III star forming halos of 3
$\times$ 10$^6 \Ms$ at redshift 20, there are $\sim$42000 halos that
have hosted a Pop III star in the volume contained within $\Delta z$,
using WMAP3 parameters with Sheth-Tormen formalism.  Clearly the
background is uniform considering the sheer number of sources within
this optically thin volume.  Local perturbations from Pop III star
formation should only affect the timing of nearby star formation but
not the global star formation rate.

\subsection{\hh~Cooling within a UVB}

Figure \ref{fig:h2form} shows SimB-LW20 twenty million years before
the central core condenses.  At this time, the core is just beginning
to cool by \hh, catalyzed by the free electrons created in the central
shocks.  In these shocks, temperatures reach $1.4 \times 10^4$ K and
electron fractions up to $10^{-3}$ exist there.  These conditions
result in \hh~formation timescales less than 25 kyr, which is
necessary to cool in a UVB of \jj~$\sim$ 1.  Within the central 10 pc,
hot and cold gas phases exist.  The hot phase exists behind the shocks
that have lower densities around 10\cubecm~and $t_{\rm{H_2}} <$ 25
kyr.  This is where \hh~cooling is catalyzed by collisional ionization
in these shocks.  The cold phase has already cooled through \hh~and
has high densities and larger $t_{\rm{H_2}}$ values.  Both phases are
apparent in the panels of Figure \ref{fig:h2form}.  Similar conditions
create \hh~in the collapses in the ``noe-'' calculations, which have
sufficient gravitational potential energy, resulting in temperatures
above $10^4$ K in central shocks.  Hence \hh~formation is possible in
the centers of high-redshift halos with virial temperatures below
10$^4$ K, even with a UVB of intensity $\jlw \sim$ 1, larger than
expected from semi-analytic models of reionization.



\subsection{Impact on Semi-analytic Models}

Two consequences of a lower critical \lya~cooling halo mass are more
frequent and earlier galaxy formation and higher mass fractions in
cooling halos.  At redshift 20, e.g., abundances of \tvir~= 4000 K
halos are an order of magnitude larger than \tvir~= 10$^4$ K halos,
resulting from the exponential nature of Press-Schechter formalism.
The mass fraction contained in these halos is three times higher than
$10^4$ K halos.  In semi-analytic models of reionization and chemical
enrichment, the star formation rate (SFR) is linearly dependent on the
collapsed mass fraction since the SFR is usually a product of mass
fraction and star formation efficiency, which is the fraction of gas
collapsing into stars \citep[e.g.][]{Haiman97a}.  The star formation
efficiency for primordial stars is $\sim10^{-3}$ with a single massive
star forming in dark matter halos with mass $\sim10^6 \Ms$
\citep{Abel02, Bromm02, Yoshida06}.  This fraction may rise to a few
percent in dwarf galaxies as widespread star formation occurs
\citep{Taylor99, Gnedin00, Walter01}.  Various studies predict that a
majority of the reionizing flux originates from dwarf galaxies
\citep[e.g.][]{Cen03, Sokasian04, Haiman06}.  If the mass contained in
star forming halos is three times greater than previously thought,
some of the predicted attributes, e.g. photon escape fractions and
star formation efficiencies, of high-redshift dwarf galaxy will
require appropriate adjustments to match observations, such as the
WMAP3 measurement of optical depth to electron scattering
\citep{Page06} and Gunn-Peterson troughs at $z \sim 6$
\citep{Becker01, Fan02}.

\section{SUMMARY}

We conducted a suite of fourteen cosmology AMR simulations that focus
on the importance of \hh~cooling with various degrees of negative
feedback.  We summarize the findings of each model below.

\medskip

1. The calculations with a UVB of \flw~= (0, $10^{-22}$, $10^{-21}$)
agree with the results of MBA01, where the critical collapse halo mass
increases as a function of UVB intensity.  

2. Above \flw~= $10^{-21}$, it had been argued that an \hh~dissociating
background would inhibit any \hh~formation until the halo could cool
through \lya~cooling.  We showed that central shocks provide
sufficient free electrons from collisional ionization to drive
\hh~formation faster than dissociation rates even in a \flw~=
$10^{-20}$ background.

3. In our ``noe-'' models, we explored when collisional ionization
becomes important and conducive for \hh~formation.  This occurs at
\tvir~$\sim$ 4000 K.  Recent major mergers above this mass scale
create complex cooling structures, unlike the non-fragmented central
cores in smaller halos.

4. Even our most extreme assumptions of \jj~= 1 (\flw~$\simeq
10^{-20}$) and no residual free electrons cannot defeat the importance
of \hh~cooling in the early universe.  

\medskip

\citet{OShea07} independently studied halo collapses with \enzo~and
similarly considered primordial gas chemistry and nine different UVB
intensities ranging from zero to \jj~= 1.  They agree with our
conclusions in that primordial gas in $\tvir < 10^4 K$ halos can
catastrophically cool and collapse even in models with $\jj \ge 0.1$.
They attribute the collapse to the increased \hh~cooling rates at
higher temperatures that is caused by greater dynamical heating in
halos with $\mvir \gsim 10^7 \Ms$.  The cooling rate per molecule is
100 times larger at 2000 K than at 500 K, typical of Pop III
star-forming halos without an UVB.  Most likely, the combination of
the elevated \hh~cooling rates and electron fractions from internal
protogalactic shocks instigate the halo collapses in a strong UVB
($\jj \ge 0.1$).

In any case, \hh~cooling triggers collapses in halos with virial
temperatures well below 10$^4$ K.  The lower critical halo mass,
corresponding to \tvir~$\sim$ 4000 K, increases mass fraction
contained in these halos by three times at redshift 20 and the number
density of high-redshift star forming halos by an order of magnitude!
By considering additional cases of extremely large negative feedback,
we have strengthened the results of MBA01 that \hh~cooling plays a key
role in high-redshift structure formation.  We conclude that a UVB
only delays and never completely suppresses \hh~formation and cooling
and subsequent star formation in these low-mass halos.

\acknowledgments{

  This work was supported by NSF CAREER award AST-0239709 from the
  National Science Foundation.  We appreciate the helpful feedback,
  which enhanced the presentation of this paper, from the referee,
  Simon Glover.  We thank Marcelo Alvarez, Greg Bryan, and Naoki
  Yoshida for providing constuctive comments on an early draft.  We
  are grateful for the continuous support from the computational team
  at SLAC.  We performed these calculations on 16 processors of a SGI
  Altix 3700 Bx2 at KIPAC at Stanford University.

}

{}

\end{document}